\documentclass[amsmath,amssymb,twocolumn,amsfonts,superscriptaddress]{revtex4-1}
\usepackage{graphicx}
\usepackage{braket}
\usepackage{stmaryrd,mathtools}
\usepackage{color}
\def \beq {\begin{equation}}
\def \eeq {\end{equation}}

\begin{document}

\title{Quantum Biometrics with Retinal Photon Counting}
\author{M. Loulakis}
\email{loulakis@math.ntua.gr}
\affiliation{School of Applied Mathematical and Physical Sciences, National Technical University of Athens, 15780 Athens, Greece}
\author{G. Blatsios}
\affiliation{Department of Ophthalmology, Medical University of Innsbruck, A-6020 Innsbruck, Austria}
\author{C. S. Vrettou}
\affiliation{First Department of Critical Care Medicine and Pulmonary Services, National and Kapodistrian University of Athens Medical School, Evaggelismos General Hospital, Athens, Greece}
\author{I. K. Kominis}
\email{ikominis@physics.uoc.gr}
\affiliation{Department of Physics and Institute of Theoretical and Computational Physics, University of Crete, 71003 Heraklion, Greece}

\begin{abstract}
It is known that the eye's scotopic photodetectors, rhodopsin molecules and their associated phototransduction mechanism leading to light perception, are efficient single photon counters. We here use the photon counting principles of human rod vision to propose a secure quantum biometric identification based on the quantum-statistical properties of retinal photon detection. The photon path along the human eye until its detection by rod cells is modeled as a filter having a specific transmission coefficient. Precisely determining its value from the photodetection statistics registered by the conscious observer is a quantum parameter estimation problem that leads to a quantum secure identification method. The probabilities for false positive and false negative identification of this biometric technique can readily approach $10^{-10}$ and $10^{-4}$, respectively. The security of the biometric method can be further quantified by the physics of quantum measurements. An impostor must be able to perform quantum thermometry and quantum magnetometry with energy resolution better than $10^{-9}\hbar$, in order to foil the device by non-invasively monitoring the biometric activity of a user.
\end{abstract}
\maketitle 
\section{Introduction}
In recent years there is an increasing need for secure biometric identification. Besides the traditional fingerprinting, the most advanced methods currently appear to be the retina and iris scan. For example, the distinctiveness of the acquired retinal image is due to the subject-specific formation of blood vessels on the retina's surface. However, all existing methods \cite{BIOMreview} are "classical", meaning that in principle they can be foiled, or equivalently, (i) there is no law of physics prohibiting such a foil, and (ii) their security is neither guaranteed nor quantified by any fundamental law of physics, but relies on the hope that the majority of potential impostors lack the means to foil them.

We here propose and analyze a biometric method based on the "single-photon" detection ability of the human retina. The method relies on the quantum estimation of $\alpha$, a parameter describing the optical/detection losses along particular optical paths ending on the retina. The estimation follows from knowledge of the incident number of photons and from the subject's response regarding the perception or not of a series of light flashes. 

The proposed method can be termed "quantum" for three reasons. First, it is based on detecting coherent light pulses containing a few tens of photons by the conscious subject supposed to be positively identified (call her Alice), (ii) an impostor pretending to be Alice, call her Eve, is forced by the physics and methodology of this biometric technique to reply randomly to the biometric device's interrogations, no matter how technologically advanced she is, in particular, no matter how good photon detectors she is equipped with, and (iii) in the event that Eve is attempting by physical means to non-invasively infer Alice's particular biometric characteristics while Alice is being interrogated by the biometric device, the ability to do so can be quantified in the context of energy resolution of quantum measurements. 

In Sec. II we introduce the basic photon statistics of low-intensity light perception by the retina and define the parameter $\alpha$ that is to be estimated and used as the biometric quantifier. In Sec. III we introduce two central performance metrics, the false-positive and false-negative identification probabilities, as well as some basic features of the biometric methodology. In Sec. IV we briefly comment on a naive biometric strategy, the quantum estimation of the $\alpha$-map across the retina, which requires an impractically long interrogation time. In Sec. V we introduce 
the central idea of the method, the random illumination of either low-$\alpha$ or high-$\alpha$ spots, which is further refined by considering a Bayesian update of the identification probability conditioned on the running record of responses. This approach leads to a realistically short interrogation time, which is further reduced by introducing in Sec. VI a "parallel" method based on pattern recognition. With conservative assumptions we obtain unprecedented performance metrics, in particular a false positive and false negative probability at the level of $10^{-10}$ and $10^{-4}$, respectively. In principle, both metrics can be readily reduced even further. Finally, in Sec. VII we examine whether the proposed method can be indeed termed "quantum". We then analyze the physical means by which an impostor could non-invasively monitor the biometric activity of Alice, infer her biometric characteristics and then pass the test. The quantum technology the impostor must possess in order to do so amounts to performing quantum thermometry and magnetometry with an energy resolution better than $10^{-9}\hbar$. 
\section{Quantum parameter estimation of the eye's "transmissivity"}
Quantum information science has turned the counterintuitive traits of quantum physics into potentially useful technology, with examples like quantum communications \cite{Wootters,Zeilinger,Patel} and cryptography \cite{Bennett,Ekert,Gisin} already leading to commercial applications \cite{Qiu}. In the case of quantum cryptography, security is guaranteed by the laws of quantum physics. Coherent sources with few photons \cite{qkd1,qkd2} and non-ideal photodetectors feature prominently in such protocols. 

Part of the recent quest to explore the possibilities biological systems offer for novel quantum technology realizations \cite{kominis_review,biochem_magn,Bowen} has been the question of how well biological photodetectors, in particular rhodopsin molecules in retinal rod cells \cite{Dowling}, compare with modern photodetectors \cite{Hadfield}. 
Historically, the role of photon statistics in vision was addressed in the 1930's \cite{Vavilov} and elucidated with human behavioral experiments in the 1940's \cite{Hecht,deVries,Rose}, single cell responses have been recorded since the 1970's \cite{Baylor,Rieke1,Rieke2,Rieke3,Rieke4}, while it was only recently that the quantum physics of human vision \cite{Nelson} was addressed with the modern experimental \cite{GisinPRL,Kr1,Kr2,Kr3,Vaziri} and theoretical \cite{GisinPRA,Horn,Pizzi, Dodel} tools of quantum optics. 
\begin{figure}[t]
\begin{center}
\includegraphics[width=8. cm]{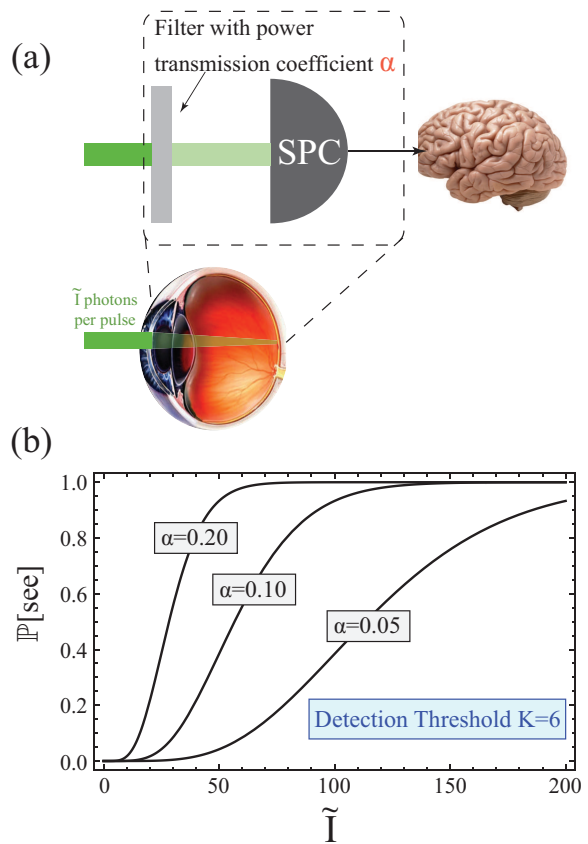}
\caption{(a) Equivalence of the proposed biometric measurement with an optical setup consisting of a filter, the transmission coefficient, $\alpha$, of which is to be estimated, followed by an ideal (unit quantum efficiency) single photon counter (SPC), having 
a threshold equal to the brain's light perception threshold of at least $K=6$ photons illuminating a ganglion receptive field. The output of the SPC is then registered by the conscious observer. The transmissivity of the filter captures the photon losses along the particular beam path ending on a particular retinal spot, including the detection efficiency of the illuminated rods. (b) Probability to see the flash of coherent light having average photon number $\tilde{I}$ for three indicative values of the parameter $\alpha$.}
\label{psee}
\end{center}
\end{figure}

It is now established that fluctuations in the perception of light by a small number of rods are governed by photon statistics. Indeed, the probability to see a flash of light of incident intensity $I$, illuminating a small area of the retina (about 0.1 ${\rm mm}^2$) for a time $\tau$ is given by the probability to count a number of photons, $n$, larger than the visual system's perception threshold, $K$. This probability, call it $\mathbb{P}[\rm see]$, is a function of $\tilde{I}$, where $\tilde{I}=I\tau$ is the average incident photon number. The actual number of detected photons, $n$,  follows \cite{Bialek} a Poisson distribution having average $\overline{n}=\alpha I\tau$. As succinctly noted in \cite{Bialek}, the parameter $\alpha$ "includes all messy details of what happens to light on its way through the eyeball". In particular, the parameter $\alpha$ includes (a) the optical losses along the beam's path to the retina, i.e. the cornea, anterior chamber, pupil, lens and vitreous body, as well as (b) the absorption probability of the particular spot of the retina being illuminated, determined by the local surface density of rod cells and their quantum efficiency. Typical values of $\alpha$ range from zero up to 0.2 \cite{Goodeve,Barlow_1977,Rieke_2005}. Although each human subject had a different $\alpha$, the data of the early experiment \cite{Hecht}, asking several individuals to respond positively or negatively on the perception of weak light flashes, could be fit by $\mathbb{P}[{\rm see}]=\sum\limits_{n=K}^{\infty}p(n;\alpha\tilde{I})$, where $p(n;\overline{n})=e^{-\overline{n}}\overline{n}^n/n!$, using a common threshold $K$.  The universal quantum noise properties of visual perception were thus revealed.

We here turn the previous arguments around and view $\mathbb{P}[\rm see]$ from the perspective of quantum parameter estimation \cite{Teklu,Escher1,Escher2}, aiming to measure the unknown, and subject-dependent parameter $\alpha$. 
In other words, while previous studies aimed at establishing single-photon detection capability and threshold, treating the subject variation of $\alpha$ as an unwanted nuisance, we take the former for granted and aim to measure $\alpha$. 

As shown in Fig. 1a, we model the particular light beam path through the eye {\em including} its detection by the rod cells as a filter, the power transmission coefficient of which, $\alpha$, we wish to estimate. The filter is followed by a single-photon counter, the output of which is registered by the conscious human brain. To estimate $\alpha$ we use (i) a coherent laser beam of known average photon number per pulse, and (ii) a conscious observer with a given light perception threshold. In other words, knowing the incident photon number and the see/don't see responses of the tested subject, we estimate $\mathbb{P}[\rm see]$, and from the dependence of $\mathbb{P}[\rm see]$ on $\alpha$ we infer $\alpha$. Examples of this dependence are shown in Fig. 1b. Interestingly, the results of the quantum measurements leading to the estimation of $\alpha$ are registered by Alice's brain, offering a first layer of security practically difficult to bypass.

The proposed biometric identification method rests on two facts \cite{perimetryBOOK}. Firstly, the values of $\alpha$ vary significantly across the retinal surface of an individual, at the level of 40 dB, and (ii) for a geometrically similar spot on the retina they vary significantly between individuals, at the level of at least 3 dB. Hence a precisely measured map of $\alpha$ values can uniquely distinguish human subjects. 
\section{Biometric performance metrics and methodology}
We here introduce two central metrics for the performance of the biometric method, the false-positive and false-negative probabilities, $p_{\rm fp}$ and $p_{\rm fn}$, respectively. The former refers to the probability that an impostor, Eve,  presents herself as Alice and successfully passes the test. The latter is the probability that Alice fails the test. The required specifications for $p_{\rm fp}$ and $p_{\rm fn}$ will determine the interrogation time. 

Before proceeding with the analysis of the various identification strategies, we make a few introductory remarks.  
Consider a single spot on the retina being illuminated with a coherent pulse of light having average photon number $\tilde{I}$ when incident on the eyeball. The probability to see the flash depends on the eye's "transmissivity" $\alpha$ corresponding to the particular light-path ending on this particular spot. In order (i) to be able to serve multiple users, and (ii) to suppress the probability that Eve foils the device, the test will rely on a whole map of retinal spots, each with its own $\alpha$ value.
We call this the $\alpha$-map. 

All strategies to be presented have a common feature: for the user to be positively identified, Alice, the device has once measured and stored the user's $\alpha$-map. The methodology for this measurement is the following. Considering a given retinal spot with a given $\alpha$ parameter, the probability to see the flash of coherent light having $\tilde{I}$ photons on average is $\mathbb{P}[{\rm see}]=\sum\limits_{n=K}^{\infty}p(n;\alpha\tilde{I})$. The sum can be evaluated in terms of the incomplete Gamma function
\beq
G_{K}(x)=\int\limits_{0}^{x}dt\ \frac{t^{K-1}e^{-t}}{(K-1)!}\label{GK},
\eeq
and is equal to $G_{K}(\alpha\tilde{I})$ (see Appendix A). Repeating the measurement for several values of $\tilde{I}$, we can use $G_{K}(x)$ to estimate $\alpha$ for that particular retinal spot. The same procedure is used for all other spots constituting the user's $\alpha$-map, which can thus be measured and stored. Obviously, the users' $\alpha$-map should not be public or accessible information. This is a common requirement of all current biometric modalities, e.g. if the database containing fingerprints or retinal images is compromised, the particular identification method is prone to failure. We further comment on this point in the closing remarks of Sec. VIII.

When Alice presents herself and asks to be identified, the device uses the information on Alice's $\alpha$-map and some identification strategy to make an inference. We next present several identification strategies in order of decreasing interrogation time, which besides the aforementioned performance metrics, is also a central parameter relevant to the practicality of the method.
\section{Biometric strategy I: Estimation of $\alpha$-map}
For this strategy, when the user Alice registers herself at the device for the first time, the value of $\alpha$ is measured and stored for each retinal spot as described before. Then, when Alice presents herself and asks to be identified, the value of $\alpha$ is estimated again using the same procedure and the result is compared with the stored values. 

The test of a single spot is considered "passed" if the estimated $\alpha$ of the subject presenting herself as Alice is within a given range of Alice's stored $\alpha$. The estimate follows after a number $\nu$ of see/don't see interrogations illuminating the particular spot. Then, the subject is positively identified as Alice if she "passes" the test for at least $\mu$ different retinal spots. Thus, the total number of see/don't see interrogations is $\nu\mu$. 
As we prove in Appendix B, in order to achieve performance metrics $p_{\rm fn}<10^{-4}$ and $p_{\rm fp}<10^{-10}$, the total number of required interrogations would be $\nu\mu\approx 2500$. That is, not only in the registration process, but also every time Alice wishes to be identified, such a large number of interrogations would be necessary, leading to a time consuming and impractical test.
\section{Biometric strategy II: Serial illumination of high-$\alpha$ and low-$\alpha$ spots}
A much better strategy is to rely on Alice's retinal spots with markedly high or low $\alpha$ values. If the device were to interrogate only low-$\alpha$ spots or only high-$\alpha$ spots, Eve would eventually learn this interrogation strategy and tune her responses appropriately. Thus, the first  central idea of this strategy is that in each interrogation the biometric device will randomly (e.g. using a quantum random number generator \cite{qrng,qrng2}) choose to illuminate a low-$\alpha$ or a high-$\alpha$ spot. Eve is unaware of what type of spot is illuminated. The second central idea is that for both low-$\alpha$ and high-$\alpha$ spots, the laser pulses will contain on average the same number of photons. Hence, even if Eve is equipped with an ideal photodetector, she can not extract any useful information from measuring the photon number in each interrogation pulse. Not knowing what type of retinal spot is being illuminated in each interrogation pulse, Eve is forced to respond randomly on perceiving or not the light flash. Moreover, no matter how many times Eve takes the test, she is always facing the same average number of photons per pulse, and hence she is always forced to respond randomly, i.e. there is no information to be acquired by Eve with time. 
These ideas underlie all three identification strategies to be presented next.

Low-$\alpha$ and high-$\alpha$ spots are defined by $\alpha_{\rm min}\leq\alpha\leq\alpha_L$ and $\alpha_H\leq\alpha\leq\alpha_{\rm max}$, respectively. As mentioned in Sec. II, it typically is $\alpha_{\rm max}\approx 0.2$ \cite{Goodeve,Barlow_1977,Rieke_2005}, while $\alpha$ varies by at least 40 dB across the retina. As will be detailed in the following, the smaller $a_{\rm min}$ the better for the proposed identification strategy. To be conservative and thus underestimate the performance of the strategy, we will consider throughout this work a 20 dB span of $\alpha$, and thus take $\alpha_{\rm min}=0.02$ and $\alpha_{\rm max}=0.2$.

For the following strategies, we assume that Alice's $\alpha$-map has been measured and stored as classical information. There are several statistical approaches to this first measurement (e.g. absolute estimate versus classification into low-$\alpha$ or high-$\alpha$), which will be explored in detail elsewhere. Now, when Alice presents herself and asks to be identified, the biometric device can resort to the following identification protocols. 
\subsection{Collective analysis of responses}
For a flash of coherent light having $\tilde{I}$ photons on average, the probabilities that Alice sees the flash in the low-$\alpha$ illumination and does not see the flash in the high-$\alpha$ illumination are bounded above by $G_K(\alpha_L\tilde{I})$ and $1-G_K(\alpha_H\tilde{I})$, respectively. The average photon number per pulse, $\tilde{I}$, is chosen so that
\begin{equation}
G_K(\alpha_L\tilde{I})=1-G_K(\alpha_H\tilde{I})=:q.
\label{qdef}
\end{equation}
This gives
\[
\frac{\alpha_H}{\alpha_L}=\frac{G_K^{-1}(1-q)}{G_K^{-1}(q)}.
\]
For example, choosing $\alpha_H/\alpha_L=3$ and setting $K=6$, the preceding equation can be solved numerically and determines the value of $q\simeq 0.1$. We can go back to \eqref{qdef} to determine $\tilde{I}$, according to Alice's $\alpha_L$ and $\alpha_H$ values. For instance, if $\alpha_L=0.05$, then $\tilde{I}\simeq 62$. 

The parameter $q$ represents the probability that Alice responds wrongly, i.e. she perceives the light pulse if a low-$\alpha$ spot is being interrogated, or she does not see the light pulse if a high-$\alpha$ spot is interrogated. The smallness of $q$ reflects the advantage Alice has over Eve, who necessarily responds randomly. To reduce $q$ we can reduce $\alpha_{L}$ and/or increase $\alpha_{H}$. On a practical level, this should not be done in a way that reveals to Eve what kind of spot is illuminated. That is, on average $\alpha$ is reduced in the periphery of the retina compared to the center. But one should not take advantage of this reduction to suppress the choice of $\alpha_{L}$, because the spatial distribution of the illuminated spots would reveal their character. Instead, one should locate neighboring spots with the highest ratio $\alpha_H/\alpha_L$.

We now consider a series of $N$ such interrogations. For $i=1,2,\ldots,N$ we define the Bernoulli random variable ${\cal X}_i$. If in the $i$-th interrogation we illuminated a low-$\alpha$ (high-$\alpha$) spot, then ${\cal X}_i=0$ (${\cal X}_i=1$) if the tested subject did not, and ${\cal X}_i=1$ (${\cal X}_i=0$) if the tested subject did see the flash. That is, $\sum\limits_{i=1}^N{\cal X}_i$ counts the wrong responses. The tested subject is identified as Alice, if $\sum\limits_{i=1}^N{\cal X}_i<Nw$ for some $w\in (q,\frac{1}{2})$ that will be determined later.

We use the same average photon number in all pulses, hence when Alice is tested, we have $\mathbb{P}_A\big[{\cal X}_i=1\big]\leq q$, for all $i=1,2,\ldots,N$. By Lemma 4.7.2 in \cite{Ash}, the probability that Alice fails the test is
\begin{equation}
p_{\rm fn}=\mathbb{P}_A\left[\sum_{i=1}^N{\cal X}_i\geq Nw\right]\leq e^{-N\mathbb{H}(w|q)},
\label{fnprob}
\end{equation}
where $\mathbb{H}(x|y)=x{\rm log}(\frac{x}{y})+(1-x){\rm log}(\frac{1-x}{1-y})$ is the relative Shannon entropy. 

On the other hand, the probability that Eve guesses wrong is $\mathbb{P}_E\big[{\cal X}_i=1\big]=\frac{1}{2}$, for $i=1,2,\ldots,N$. Hence, the probability that Eve passes the test is
\begin{equation}
p_{\rm fp}=\mathbb{P}_E\left[\sum_{i=1}^N{\cal X}_i< Nw\right]\leq e^{-N\mathbb{H}\left(w |\frac{1}{2}\right)}.
\label{fpprob}
\end{equation}
Now, since $w\in (q,\frac{1}{2})$, it follows that $\mathbb{H}\big(w\,\big|\,q)$ appearing in \eqref{fnprob} is an increasing function of $w$, whereas $\mathbb{H}\big(w\,\big|\,{1\over 2})$ appearing in \eqref{fpprob} is a decreasing function of $w$.
Hence the respective bounds for $N$ obtained from \eqref{fnprob} and \eqref{fpprob} have the opposite $w$-dependence, and to minimize $N$ we set 
\[
\log \big(\frac{1}{p_{\rm fn}}) \mathbb{H}\big(w\,\big|\,\frac{1}{2}\big)=\log \big(\frac{1}{p_{\rm fp}}) \mathbb{H}\big(w\,\big|\,q\big),
\]
solving which we obtain $w$. Inserting this into either  \eqref{fnprob} or \eqref{fpprob} we then obtain $N$. For instance, requiring $p_{\rm fn}=10^{-4}$ and $p_{\rm fp}=10^{-10}$, we get $w\simeq 0.22$ and $N=138$ interrogations, which is an order of magnitude lower than the "naive" strategy of the previous section.
\subsection{Real-time Bayesian update of identification probability} 
We will now demonstrate that we can significantly speed up the process, by more than a factor of 2, terminating the test when the conditional probability that the tested subject is Alice, given her running record of answers, reaches a satisfactory level. 

In particular, let $\mathbb{P}[A]\in(0,1)$ be the a priori probability that the tested subject is Alice and for $i\in\mathbb{N}$ denote by 
$
\mathbb{P}\big[A\,|\,{\cal F}_i\big]
$
the conditional probability that the tested subject is Alice, given her answers to the first $i$ interrogations. We also define $S_i=1$ ($0$), if the response of the tested subject to the $i$-th interrogation is {\em see} ({\em no see}). Based on Bayes's rule, we can update the conditional probabilities $\mathbb{P}\big[A\,|\,{\cal F}_i\big]$. For instance, if $i=1$ and $v\in\{0,1\}$, then the conditional probability $\mathbb{P}\big[A\big | \alpha_1=\alpha, S_1=v\big]$ is equal to
\beq
\frac{\mathbb{P}\big[S_1=v\big| A, \alpha_1=\alpha\big]\mathbb{P}[A]}{\sum\limits_{X=A,E}\mathbb{P}\big[S_1=v\big| X, \alpha_1=\alpha\big]\mathbb{P}[X]}.\label{cond}
\eeq
Here $\mathbb{P}[E]=1-\mathbb{P}[A]$ is the a priori probability that the tested subject presenting herself as Alice is not Alice, and $\mathbb{P}\big[S_1=v\big| X, \alpha_1=\alpha\big]$ is the conditional probability that the response is $v$, given the subject is X (where X=A,E) and the tested spot has $\alpha_1=\alpha$. 

For iterating the calculation it is helpful to define $Z_A(\alpha,v)=\mathbb{P}\big[S_1=v\big| A, \alpha_1=\alpha\big]$. We can write
\begin{align*}
Z_A(\alpha,v)&=\begin{cases} G_K(\alpha\tilde{I}),&\text{for }v=1\\1-G_K(\alpha\tilde{I}),&\text{for }v=0.\end{cases}
\end{align*}
To calculate \eqref{cond}, we need to set $\mathbb{P}\big[S_1=v\big| E, \alpha_1=\alpha\big]$. The choice of this parameter is made by the test designer and reflects the designer's belief to get the answer $S_1=v$, given that the subject who claims to be Alice is not. Intuitively, we could set this to 1/2, given that Eve can not guess what type of spot (low-$\alpha$ or high-$\alpha$) the device illuminates. Another reasonable choice is to set this parameter equal to $p=\mathbb{E}_{\alpha}\big[G_K(\alpha\tilde{I})\big]$. After all, this is the best guess one can make for the probability that Alice sees a flash without any information on the value of $\alpha$. It is readily shown (see Appendix C) that $p\in (\frac{1-q}{2},\frac{1+q}{2})$. 

With this choice we can get analytically tractable and nearly optimal bounds for the identification thresholds and interrogation time, regardless of the answering strategy a potential impostor may follow. At the end, the judicious choice of $p$ is reflected in the success and optimality of the identification algorithm.
\begin{figure*}[th]
\begin{center}
\includegraphics[width=17 cm]{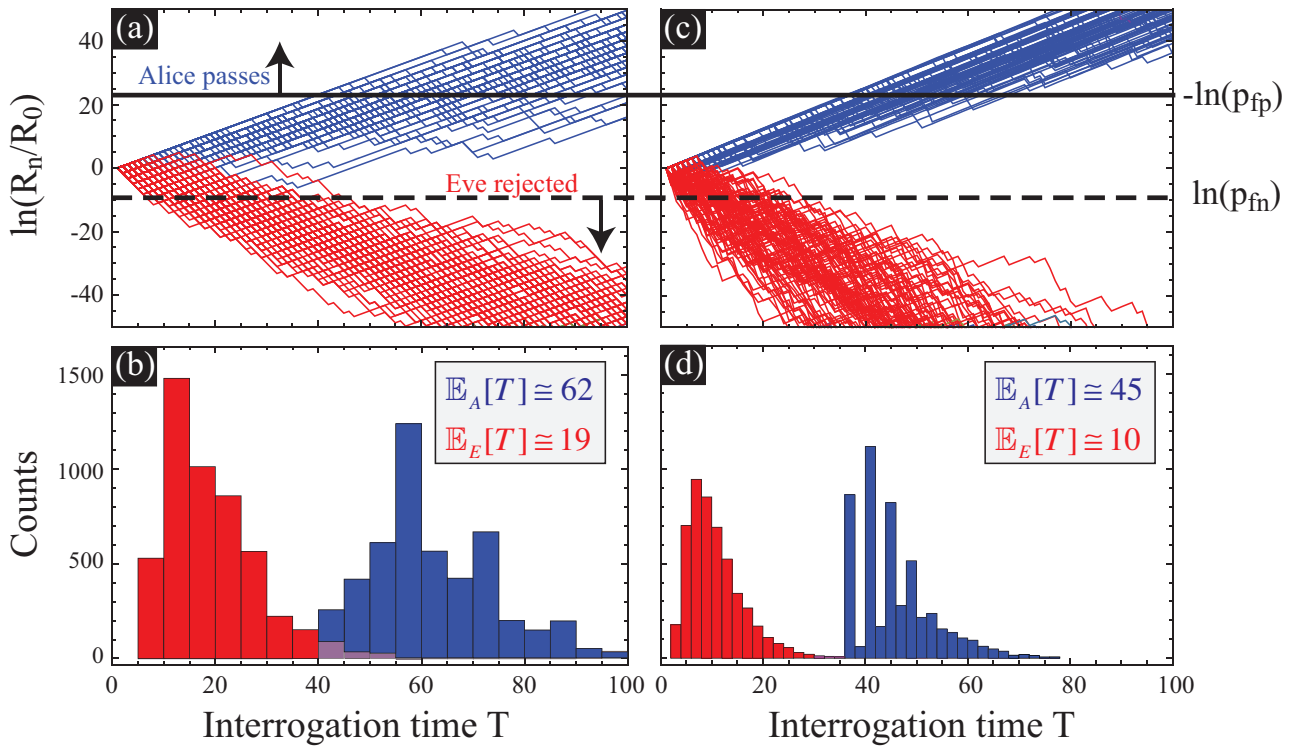}
\caption{Random walks of ${\rm ln}(R_n/R_0)$ (Alice blue and upwards, Eve red and downwards) and distribution of interrogation times $T$. (a,b) Illumination of spots with (a) $\alpha_{L}=0.05$ and $\alpha_{H}=0.15$. For these values of $\alpha_{L}$ and $\alpha_{H}$ it is $q=0.096$ and $\tilde{I}=62.4$. (c,d) Illumination of spots with low-$\alpha$ and high-$\alpha$ uniformly distributed in [0.02,$\alpha_{L}$] and [$\alpha_{H}$,0.18], respectively. If ${\rm ln}(R_n/R_0)$ becomes higher (lower) than the upper (lower) threshold defined by $-\ln(p_{\rm fp})$ shown by the black solid line ($\ln(p_{\rm fpn})$ shown by the dashed black line) in (a,c), the interrogation stops and Alice is identified (Eve is rejected). The obtained mean values of $T$ shown in (b) and (d) are consistent with the bounds \eqref{bounds}. In (a) and (c) we plot just 100 walks, the distributions (b) and (d) were obtained from 5000 walks.}
\label{rw}
\end{center}
\end{figure*}
To proceed, define
\beq
Z_E(p,v)=\begin{cases} p&\text{for }v=1\\1-p,&\text{for }v=0.\end{cases}
\eeq
From Bayes's rule for $i=1$ we find 
\[
\mathbb{P}\big[A\big|{\cal F}_1\big]=\frac{Z_A(\alpha_1,S_1)\mathbb{P}[A]}{Z_A(\alpha_1,S_1)\mathbb{P}[A]+Z_E(p,S_1)\mathbb{P}[E]}.
\]
Iterating this argument, it follows for any non-negative integer $i$ that 
\[
\mathbb{P}\big[A\big| {\cal F}_i\big]=\frac{Z_A(\alpha_i,S_i)\mathbb{P}\big[A\big|{\cal F}_{i-1}\big]}{Z_A(\alpha_i,S_i)\mathbb{P}\big[A\big|{\cal F}_{i-1}\big]+Z_E(p,S_i)\mathbb{P}\big[E\big|{\cal F}_{i-1}\big]}.
\]
If we define the odds ratio after $i$ interrogations
\[
R_i=\frac{\mathbb{P}\big[A\big| {\cal F}_i\big]}{1-\mathbb{P}\big[A\big|{\cal F}_{i}\big]}=\frac{\mathbb{P}\big[A\big| {\cal F}_i\big]}{\mathbb{P}\big[E\big|{\cal F}_{i}\big]},
\]
the updating rule takes the simple form $R_i=Z_A(\alpha_i,S_i)R_{i-1}/Z_E(p,S_i)$, hence
\[
R_n=R_0\prod_{i=1}^n \frac{Z_A(\alpha_i,S_i)}{Z_E(p,S_i)}.
\]
\subsubsection{Identification Thresholds}
We may now set two thresholds $x,y$, with $0<x<1<y$ and stop the interrogation at the random time $T\in\mathbb{N}$ as soon as the ratio $R_n/R_0$ falls outside the interval $(x,y)$.
The tested subject is identified as Alice, if $\frac{R_T}{R_0}\ge y$, and is rejected, if 
$\frac{R_T}{R_0}\le x.$ The thresholds $x$ and $y$ are set by the desired specifications for $p_{\rm fp}$ and $p_{\rm fn}$. 

In Appendix D we prove that the process $\{R_n^{-1}\}_{n\ge 0}$ is a martingale for Alice, and the process $\{R_n\}_{n\ge 0}$ is a martingale for Eve, regardless of her answering strategy. A stochastic process $\{M_n\}_{n\ge 0}$ is a martingale when, given the history up to any time $n$, its expected value at time $n+1$  is the same as its value at time $n$, much like a gambler's fortune in a fair game. That is, the definition of a martingale is 
\[
\mathbb{E}\big[{M}_{n+1}\big|{\cal F}_{n}\big]=M_{n},
\]
a consequence of which is that the martingale's expectation is constant in time. Notably, this expectation does not change even if we stop the martingale with a random strategy such as the one discussed here. 
This is the optional stopping theorem (cf. 10.10 in \cite{williams}), which we use next to find the expected values of $R_T$ and $1/R_T$ for Eve and Alice, denoted by $\mathbb{E}_E\big[R_T\big]$ and $\mathbb{E}_A\big[\frac{1}{R_T}\big]$, respectively. For Eve we get
\begin{align*}
R_0&=\mathbb{E}_E\big[R_T\big]\ge R_0y\ \mathbb{P}_E\big[\frac{R_T}{R_0}\ge y\big].
\end{align*}
Hence, $\mathbb{P}_E\big[\frac{R_T}{R_0}\ge y\big]\le \frac{1}{y}$, and it suffices to take $y=\frac{1}{p_{\rm fp}}$ to achieve the desired false positive probability.
The optional stopping theorem for Alice gives
\begin{align*}
\frac{1}{R_0}&=\mathbb{E}_A\big[\frac{1}{R_T}\big]\ge \frac{1}{xR_0} \mathbb{P}_A\big[\frac{R_T}{R_0}\le x\big].
\end{align*}
Hence,  $\mathbb{P}_A\big[\frac{R_T}{R_0}\le x\big]\le x$, and it suffices to take $x={p_{\rm fn}}$ to achieve the desired false negative probability.\\[2mm]
\subsubsection{Duration of Interrogation}
We finally address the duration of the test for Alice and Eve.
Note first that
\begin{equation}
\ln\left(\frac{R_n}{R_0}\right)=\sum_{i=1}^n\ln\left(\frac{Z_A(\alpha_i,S_i)}{Z_E(p,S_i)}\right).
\label{logodds}
\end{equation}
The test ends when $\ln(R_n/R_0)$ exits the interval $(\ln p_{\rm fn},-\ln p_{\rm fp})$ and the test taker is identified as Alice
if $\ln\big(R_T/R_0\big)\ge -\ln p_{\rm fp}$, or is rejected if $\ln\big(R_T/R_0\big)\le \ln p_{\rm fn}$. Hence, we may view the test as a random walk starting from 0, 
with increments $\ln\big({Z_A(\alpha_i,S_i)}/{Z_E(p,S_i)}\big)$. In Appendix C we establish that Alice's walk always has a drift to the right, while Eve's walk always has a drift to the left. 
As shown in Appendix E, we can use \eqref{logodds} and the optional stopping theorem to estimate the expected stopping times for Alice and Eve. They read
\begin{align}
\mathbb{E}_A\big[T\big]&\le \frac{\ln\big(\frac{2}{(1-q)p_{\rm fp}}\big)}{\mathbb{H}\big(q\,\big|\,\frac{1}{2})}\label{EA}\\
\mathbb{E}_E\big[T\big]&\le \frac{2\ln\big(\frac{2q_{\rm min}p_{\rm fn}}{1+q}\big)}{\ln\big(4q(1-q)\big)},\label{bounds}
\end{align}
where $q_{\rm min}=\min\{G_K(\alpha_{\rm min}\tilde{I}), 1-G_K(\alpha_{\rm max}\tilde{I})\}$. With our choice of $p_{\rm fp}=10^{-10}$, $p_{\rm fn}=10^{-4}$,  $q=0.1$ and $\alpha_{\min}=0.02$ we obtain $\mathbb{E}_A\big[T\big]\le 65$ and $\mathbb{E}_E\big[T\big]\le 28$. The above estimates on $T$ are rigorous bounds and they are sharp if we only target retinal spots with $\alpha=\alpha_L$ or $\alpha=\alpha_H$ \cite{note}.

A Monte Carlo simulation for the random walk of $\ln(R_n/R_0)$ is shown in Fig.\ref{rw}. In Figs.\ref{rw}(a,b) we consider the simple scenario of targeting spots with $\alpha=\alpha_L$ or $\alpha=\alpha_H$. Targeting spots with $\alpha<\alpha_L$ or $\alpha>\alpha_H$ improves the running time of the algorithm. This is shown in Figs.\ref{rw}(c,d), where we use a uniform distribution of $\alpha$ in [0.02,0.05] for low-$\alpha$, or in [0.15,0.18] for high-$\alpha$.

It is seen that with a modest number of about 50 interrogations (even fewer if we can increase $\alpha_{H}$ and decrease $\alpha_{L}$) we can identify Alice and meet the desired specifications for $p_{\rm fp}$ and $p_{\rm fn}$. This interrogation time is two orders of magnitude smaller than the time needed to absolutely estimate the $\alpha$-map as discussed in Sec. IV
\subsubsection{Optimality of the Algorithm}
For practical reasons, identifying Alice with the shortest possible test duration is obviously of interest. We here evaluate the optimality of the previous identification algorithm, in particular the bound \eqref{EA}. To do so, we find the lower bound of interrogations $N$ needed to achieve a desired $p_{\rm fp}$ in a series of questions having binary answers, given the probability of an incorrect answer by the subject supposed to pass the test. This bound is just dependent on binomial statistics and thus is general, i.e. independent of any particular context and physical realization of the questions. 

If $q$ is the probability of an incorrect answer by Alice, the number of incorrect answers in $N$ interrogations follows the binomial distribution, the median of which lies in $\big[[qN], [qN]+1\big)$, where $[x]$ is the integer part of $x$. By the definition of the median, the probability that at least $[qN]$ answers are incorrect is at least 1/2. To avoid rejecting Alice as often as half of the times she is tested, the tested subject has to be identified as Alice if she gives at most $[qN]$ incorrect answers.

Now suppose Eve answers randomly. $N$ has to be large enough so that the event $E_q$, that Eve gives at most $[qN]$ incorrect answers, has probability smaller than $p_{\rm fp}$. Eve's incorrect answers also follow a binomial distribution, so Lemma 4.7.2 in \cite{Ash} gives
\[
p_{\rm fp}\ge \mathbb{P}\big[E_q\big]\ge \frac{1}{\sqrt{8Nq_N(1-q_N)}}e^{-N\mathbb{H}\big(q_N\big|\frac{1}{2}\big)},
\]
where $q_N=[qN]/N$. Taking the approximation $q_N\simeq q$ this equation gives that
\[
N\mathbb{H}\big(q\big|\frac{1}{2}\big)+\frac{1}{2}\ln\big(8Nq(1-q)\big) \gtrapprox \ln \frac{1}{p_{\rm fp}}.
\] 
Comparing this lower bound for $N$ with \eqref{EA}, we see that the average number of interrogations required by the algorithm we propose is close to the absolute lower bound. For example, setting $q=0.1$ and $p_{\rm fp}=10^{-10}$, we find 
$N\geq 57$, to be compared with the upper bound of 65 found previously. 
\section{Biometric strategy III: pattern recognition}
\begin{figure*}[th!]
\begin{center}
\includegraphics[width=18 cm]{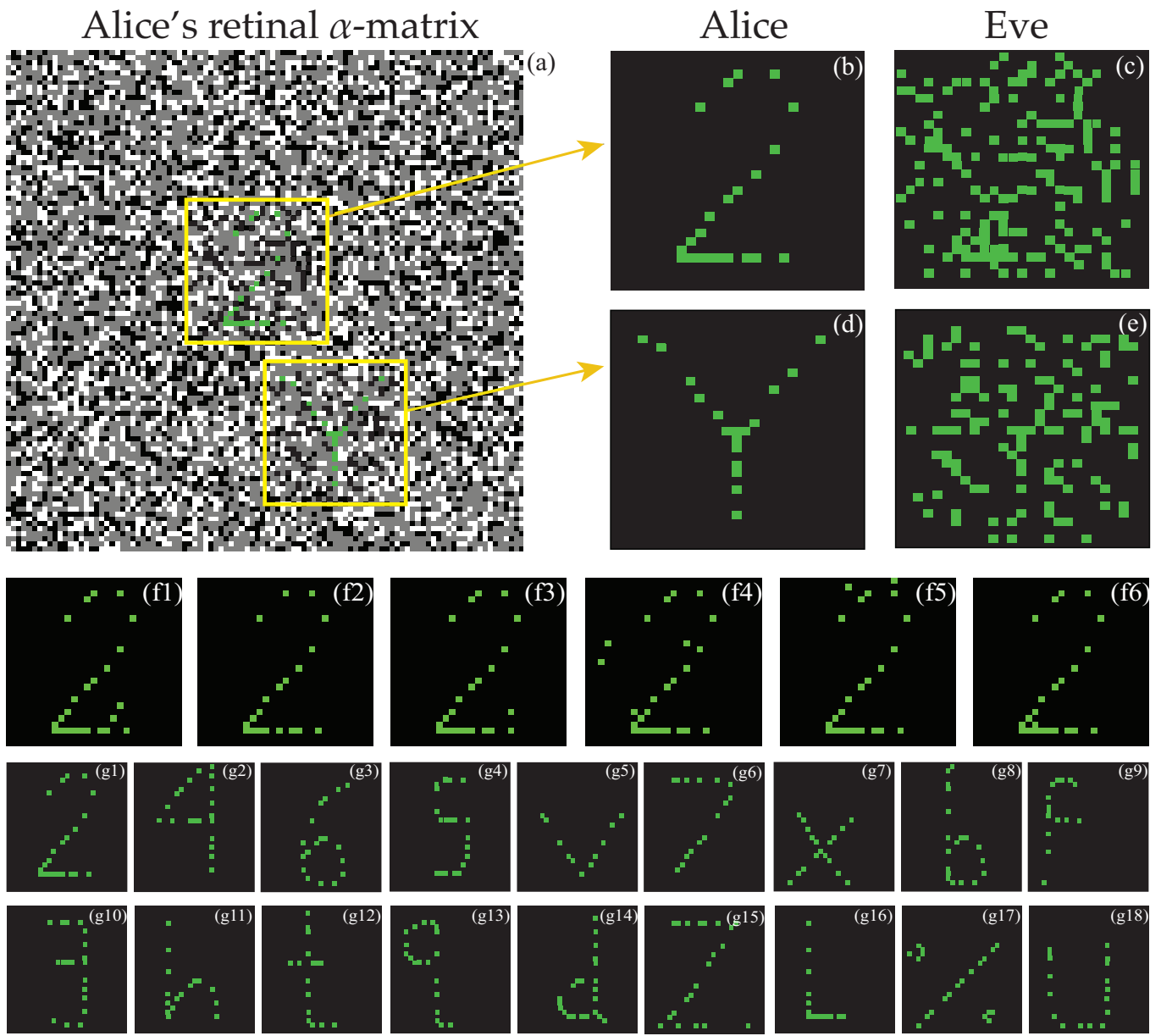}
\caption{Biometric Strategy. (a) We suppose Alice's retina consists of $100\times 100$ non-overlapping ganglion receptive cells that can be individually addressed with a laser pulse. The light-path ending in each one of those is described by a characteristic $\alpha$ value, which here takes a random value between $\alpha_{\rm min}=0.02$ and $\alpha_{\rm max}=0.2$ (we ignore the spatial dependence of $\alpha$, i.e. in a real retina the average $\alpha$ decreases going from center to periphery). The color coding in (a) is white for high-$\alpha$ values, i.e. $0.16\leq\alpha\leq 0.2$, grey for $0.04\leq\alpha\leq 0.16$ and black for low-$\alpha$ values, i.e. $0.02\leq\alpha\leq 0.04$. We can easily find a number of white pixels forming a pattern, for example the number "2" (upper yellow box) or the letter "Y" (lower yellow box). To the set of pixels forming the pattern we add another set of low-$\alpha$ pixels so that in the combined set there is no discernible pattern. However, Alice would ideally (only high-$\alpha$ pixels fire) just see the pattern, as shown in (b) and (d). In contrast, even if Eve is equipped with an ideal photodetector, all she would see is the combined set of pixels, shown in (c) and (e). Realistically, Alice can miss some high-$\alpha$ pixels and see noise from some firing low-$\alpha$ pixels. A few realizations of a Poisson simulation (a pixel $(i,j)$ is lighted if a Poisson random variable with average $\alpha_{ij}\tilde{I}$ exceeds the detection threshold $K=6$, where we used $\tilde{I}=72$) are shown in (f1)-(f6) for the case of "2". (g1)-g(18) Example for realizing $p_{\rm fp}\approx 10^{-10}$ with $m=8$ interrogations. Among the illuminated spots of (c) we can form at least 18 patterns, in particular 2, 4, 6, S, v, 7, x, b, f, 3, h, t, q, d, Z, L, \%, U. The tested subject is offered these 18 patterns from which she is supposed to pick her response. While Alice would most probably perceive "2", since "2" is the pattern formed by Alice's high-$\alpha$ spots, Eve's only option would be to randomly pick her response among the 18 choices, thus after 8 interrogations one could achieve $(1/18)^8\approx 10^{-10}$ for the false-positive probability.}
\label{pattern}
\end{center}
\end{figure*}
To even further reduce test time, we finally describe a "parallel" interrogation scheme using a pattern recognition method. We will here not delve into discussing receptive fields and complex cognitive interactions among them \cite{Lindeberg, Sterling}, since we assume that we illuminate non-overlapping ganglion receptive fields, which are roughly \cite{Sterling} 0.2 mm wide, so we take a $2~{\rm cm}\times 2~{\rm cm}$ retinal surface to approximately contain 10,000 pixels.

When Alice presents herself and asks to be identified, the device picks among all possible retinal spots of Alice a small subset with high $\alpha$ in such a way that they form a particular pattern, e.g. the number "2". Moreover, the device picks another set of spots having low $\alpha$, and illuminates both the former and the latter with the same average photon number. In Fig.\ref{pattern} we describe how this method works. We first generate a $100\times 100$ matrix of retinal spots with $\alpha$ randomly picked for each spot from the interval between $\alpha_{\rm min}$ and $\alpha_{\rm max}$. This $\alpha$-matrix is shown in Fig.\ref{pattern}a, where black pixels denote $0.02=\alpha_{\rm min}\leq\alpha\leq 0.04$, grey pixels $0.04\leq\alpha\leq 0.16$ and white pixels $0.16\leq\alpha\leq 0.2=\alpha_{\rm max}$. Although the retina has a spatial $\alpha$-distribution far from random, this simulation is adequate to exhibit the features of our model which can be extended to realistic spatial distributions. In any case, the 10,000 pixels provide many possibilities for picking a small subset (containing around 100 pixels), which includes a recognizable pattern. This is not to imply that 10,000 pixels ought to be measured in Alice's retina. The simulation is just depicting that patterns can be readily found in an array of randomly varying $\alpha$.

For example, two such subsets are contained in the two (yellow) boxes of Fig.\ref{pattern}a. Among the pixels in those boxes, we choose some having high-$\alpha$ in such a way that they form the number "2" (upper box) or the letter "Y" (lower box). We augment those high-$\alpha$ pixels with a larger number of low-$\alpha$ pixels, in such a way that the combined set cannot reveal the pattern. We illuminate all with pulses having the same average number of photons $\tilde{I}$.  An ideal photodetector would see the patterns in Figs.\ref{pattern}(c,e). This is what Eve would observe in the best possible scenario she is equipped with an ideal photon detector, and she would obviously be unable to infer any pattern. In contrast, Alice would observe the patterns in Figs.\ref{pattern}(b,d) in the ideal case that the high-$\alpha$ pixels perceive light, and the low-$\alpha$ pixels do not. 

In reality, both high- and low-$\alpha$ pixels of Alice might or might not perceive the light flashes. To simulate that, we sample independent Poisson variables for each of the illuminated pixels (i,j) with average $\alpha_{ij}\tilde{I}$, so that the pixel is bright if the realization of the process exceeds the detection threshold $K$. Some of the possible realizations are shown in Figs.\ref{pattern}{(f1-f6).

Alice is interrogated $m$ times, asked to respond on which symbol she saw, and given a choice of $M$ symbols, where $M$ could easily be around 40 (alphabet, numbers, etc.). It is straightforward to calculate the false positive probability. Even if Eve is equipped with an ideal photodetector, all she measures is an apparently random pattern of illuminated spots with the same average number of photons. As she is unaware of Alice's $\alpha$-map, the only option Eve has is to respond randomly given the same choice of $M$ symbols, hence the probability that she correctly responds to $m$ questions is $p_{\rm fp}=(1/M)^m$. For $M=40$ and just $m=6$ interrogations we obtain $p_{\rm fp}\approx10^{-10}$. Practically, $m=6$ interrogations can be realized in less than one minute of test time.

An example for the realization of the above scheme is the following. In the illuminated set of spots shown in Fig.\ref{pattern}c, the tested subject could be offered the choice of $M=18$ patterns shown in Figs.\ref{pattern}g1-g18. These patterns (and perhaps more) are formed by spots which are a subset of the illuminated spots of Fig.\ref{pattern}c. While Alice would most probably perceive "2", since "2" is the pattern formed by the high-$\alpha$ spots of Alice, Eve would be clueless as to which pattern Alice would perceive and would have to randomly pick one of the 18 patterns as her response. Thus in this case, with $m=8$ interrogations we can achieve $p_{\rm fp}\approx10^{-10}$. 
\newline\newline\newline
We now calculate the probability that Alice fails the test. We assume that the pattern to be recognized is formed by $n_H$ high-$\alpha$ spots. We also illuminate $n_L$ spots of low-$\alpha$ to create noise for Eve. Suppose the probability that Alice fails to see a pattern spot is bounded by $p_H$, while the probability to see a noise spot is bounded by $p_L$. Without entering into cognitive issues of object recognition, which would only increase Alice's chances to recognize the pattern, we assume that Alice fails to recognize the pattern, if she wrongly gets at least $k>n_Hp_H$ of the pattern spots, or if she sees at least $\ell>n_Lp_L$ noise spots. Then, the probability $P$ that Alice fails to recognize the pattern satisfies
\[
P\le e^{-n_H\mathbb{H}\big(\frac{k}{n_H}|p_H\big)}+e^{-n_L\mathbb{H}\left(\frac{\ell}{n_L}|p_L\right)}.
\]
Thus the probability that Alice passes $m$ questions is larger than $(1-P)^m$, hence 
\beq
p_{\rm fn}\leq 1-(1-P)^m
\eeq 
As in the examples of Figs.\ref{pattern}(b,d) we take the pattern to be composed of $n_H\approx 25$ pixels, and add another $n_L\approx 75$ "noise" pixels, so that in total we illuminate about 100 pixels. We assume that recognition is possible if Alice misses at most 20\% of the pattern's pixels and at most an equal number of noise pixels are perceived by her, i.e. we take $k=\ell=5$. We choose the average photon number per pixel $\tilde{I}$ by minimizing $p_{\rm fn}$ and find the minimum $p_{\rm fn}=5\times 10^{-4}$ occurs at $\tilde{I}=72$. 

The achieved $p_{\rm fn}$ depends dramatically on the values of $\alpha_L$ and $\alpha_H$. Thus, if it is possible to create patterns+noise by finding spots with lower $\alpha_L$ and higher $\alpha_H$, the probability that Alice will fail the test and will have to retake it can be further reduced. Finally, we note that we refrained from introducing cognitive aspects of object recognition into the calculation of $p_{\rm fn}$, i.e. we chose as 20\% what seems to be 
a reasonable percentage of missed spots and added noise. We could have considered maximum-likelihood image classification strategies \cite{Wernick}, however the fundamental results would be unchanged.
\section{The quantum aspect of this biometric method}
There are several reasons why this biometric method can be termed "quantum". (i) The method is essentially a quantum parameter estimation process, where a classical parameter (here the optical loss of a particular light path ending on a particular retinal spot) is estimated using a 
quantum process (here photodetection). Rod cells realize the single photon detector, while the conscious brain realizes the counter. (ii) While in this work we utilize coherent light for illuminating the retinal spots, the same measurement can in principle be 
performed with single-photon sources \cite{single_photon_1,single_photon_2} or other non-classical light sources. In fact, such sources provide an advantage for the biometric methodology that will be explored elsewhere. (iii) Finally, unlike other biometric methods, the security of this method can be explicitly quantified by the laws of quantum measurements and stated in terms of energy measurement resolution given in units of $\hbar$. This is presented in detail in the following subsection.

\subsection{Quantum technology required to non-invasively access the biometric information}
As discussed previously, if an impostor, Eve, presents herself as Alice and asks to be identified by the device, no matter how technologically advanced Eve is, the only strategy she has is to respond randomly to the device's interrogations. We will now entertain a different way to foil the device.  Suppose that Eve is proximal to Alice (e.g. 1 m away), and while Alice is being interrogated by the biometric device, Eve is secretly operating a highly advanced quantum sensor monitoring Alice's activity. For example, if Eve could measure the number of photons scattered in Alice's eye, and at the same time monitor her brain activity, she might be possible to correlate these observations and infer Alice's $\alpha$-map. 

Before proceeding with the estimates, we note that in the scenario of a single retinal spot being illuminated, if Eve is aware of (or measures) the incident photon number $\tilde{I}$, then she only needs to "detect" brain activity in order to infer $\alpha$ of that spot. We can easily force her to also require sensitive thermometry by changing (randomizing) the incident photon number in each interrogation. 

Now since $\alpha\approx 0.1$ on average, there will be on the order of $n_s=50$ scattered photons for $\tilde{I}=50$ incident photons. Eve could try to non-invasively detect a minute temperature change in Alice's eye (or in turn in the environment), due to the energy deposited by the scattered photons. We use simple order-of-magnitude estimates and consider just a single illuminated pixel, i.e. we ignore the issue of spatial resolution of Eve's measurement needed for the multi-pixel scenario, and we also ignore the thermal eye-environment contact as well as the eye's physiological cooling mechanisms (eye-body thermal contact). So we will heavily underestimate the level of technology required by Eve.

Let us approximate the eye, having mass about $m=10~{\rm g}$, by water, the specific heat of which $c_w=4~{\rm J/g/C}$. Thus the $n_s$ scattered photons of wavelength $\lambda=532~{\rm nm}$ would deposit $n_s(hc/\lambda)$ of energy into the eye, so Eve should be able to resolve the eye's temperature change $\delta\theta=n_s(hc/\lambda)/mc_w\approx10^{-19}~^{\circ}{\rm K}$. This would happen during the pulse time $\tau=0.1~{\rm s}$, so Eve should be able to non-invasively measure the thermal energy deposited in Alice's eye with a resolution $(k_{B}\delta\theta)\tau\approx 10^{-9}\hbar$. 

Furthermore, it is known that the magnetic activity of the brain can be modeled with current dipoles \cite{MEG} producing on the head's surface, being about 10 cm away from the dipoles,  magnetic fields on the order of 1 fT in the case of visual perception \cite{MEG_Vision}. Hence at a distance 1 m away the fields will be attenuated (due to the $1/r^3$ distance dependence of dipolar fields) by a factor $10^3$. Eve should thus be equipped with a magnetometer having sensitivity at the $0.1~{\rm aT}/\sqrt{\rm Hz}$ level \cite{RomalisMagn,RomalisMEG}. Using the electron's gyromagnetic ratio, this translates to an energy resolution of $10^{-9}\hbar$.  While modern optical magnetometers \cite{Allred,KomNat} are about three orders of magnitude away from $10^{-9}\hbar$, quantum thermometers \cite{Th1,Th2,Th3} are many more orders of magnitude away.
\section{Discussion}
(i) The proposed biometric method resembles static visual perimetry \cite{perimetryBOOK,perimetry1,perimetry2}, a diagnostic tool in ophthalmology used to assess e.g. glaucomatous disease. Perimetry is the measurement of the differential light sensitivity, i.e. the threshold of perception of a test object projected on the visual field against its background. Yet, there is a significant difference of our approach from visual perimetry. The latter can hardly profit from the idea of interrogating only low-$\alpha$ or high-$\alpha$ spots in any of our serial, Bayesian or "parallel" pattern recognition schemes discussed in Secs. V-VI. The reason is that pathological patterns could appear in any particular retinal spot or area, and hence their detection requires the time-consuming $\alpha$-map estimate discussed in Sec. IV. In other words, detecting pathology poses a different statistical inference problem than verifying identity. 

(ii) Even in the absence of some pathology, because of the psychophysical character of the test, a fluctuation in the measured threshold at a particular testing point of the same individual at different times can be observed. This could be related to the visual perception threshold $K$, the exact value of which is still debated \cite{Vaziri,Rieke_2005}. Moreover, its value might depend on other physiological parameters and not even be constant for the same subject. We used $K=6$, but our estimates are robust and can be extended to include a distribution of $K$. Similarly, it is known that the parameter $\alpha$ is age-dependent. Again, such issues can be mitigated by proper statistical analysis or slightly increased number of required interrogations, i.e. they do not alter the fundamental principle or the underlying statistics. Another technical point is the ability to consistently target specific areas of the retina. This is accomplished by fundus based eye-tracking \cite{fundus} as applied e.g. in optical coherence tomography \cite{oct}.

(iii) We considered the perception threshold $K$ to be spatially constant across the retina. A spatial variation of $K$ could be considered to lead to a deterioration of the identification algorithm's performance. Counterintuitively, the opposite is the case.
The reason is that the algorithm is based on the variability of $\alpha$ across the retina. But essentially, it is the variability of $\mathbb{P}\big[\text{see}\big]$ that we take advantage of. Now, the probability $\mathbb{P}\big[\text{see}\big]$ is given by $G_{K}(\alpha\tilde{I})$. Tacitly assuming that $K$ is constant, the variability of $\alpha$ is translated into the variability of $\mathbb{P}\big[\text{see}\big]$. If the parameter $K$ also varies independently, it will add an independent “channel” of variability to $\mathbb{P}\big[\text{see}\big]$, and thus give the physical realization of the algorithm a greater versatility.

(iv) Our method is a scotopic measurement, i.e. in the dark-adapted eye, which is known to be tedious for the human subject under examination. The required dark adaptation (lasting at least half an hour) is a serious limitation on the practicality of the method. However, what we aim at in this work is the proof of principle of the methodology using the rod response, which is well studied and documented in terms of single-photon sensitivity. We expect that cones and the light adapted eye, which does not suffer from this limitation, will provide similar capabilities for realizing the method.  While cone single photon detection capabilities are not broadly established so far, cones have some unique characteristics, like a much shorter integration time and a much faster response/recovery time to repeated stimuli compared to rods. Furthermore they have a much wider dynamic range, making fine threshold determination easier. Thus we expect we will be able to apply the same or a similar methodology in photopic conditions.

(v) Any biometric method one can imagine suffers from the possibility that the impostor, Eve, can have access to the biometric data. For all biometric methods known so far the biometric data are stored in a computer database as classical information. If the privacy of these databases is compromised, the security of all known biometric methods can be readily thwarted. The biometric method proposed here is not an exception in this respect, i.e. it is assumed that the impostor cannot have access to the 
stored biometric data of the device's users.

(vi) Another common feature of all known biometric modalities is that they cannot resist forceful tactics by the impostor. The method proposed here is also prone to failure if forceful tactics are allowed. Regarding for example the pattern recognition strategy, Eve could force Alice to reveal which $j$ patterns she is perceiving (assuming in the worst case that not all $M$ patterns can be formed on Alice's retina, i.e. $j\ll M$), so Eve could increase here chances of success to $(1/j)^m$. This scenario can be easily mitigated by combining tests in both eyes, effectively doubling $j$ and possibly requiring slightly more interrogations than $m=6$. Essentially, the $M$ patterns to choose from in order to reply to each interrogation could be public information. 

A more extreme scenario would be that Eve acquires the biometric device, forcefully measures Alice's $\alpha$-map, and then passes the test with a properly designed photodetector mimicking Alice's eye. We here wish to entertain another scenario, because it has a scientific interest in its own right. Since Alice would still have to cooperate by properly responding to the light flash interrogations, in the event that she would not, Eve could resort to a more subtle approach and measure (with Alice either being conscious or sedated) Alice's pupillary light reflex \cite{AJO1964,Pennesi,Do,Costic,Rukmini}, hoping to extract information on Alice's $\alpha$-map. However, existing evidence suggest a higher detection threshold for the pupillary reflex, and in general, the relation of the physiological backgrounds of light perception and pupillary reflex is poorly understood. In any case, the relevance of this reflex measurement to the proposed method, and its own potential for yet another biometric quantifier will be addressed elsewhere.

Summarizing, we have here presented the principal workings of a quantum optical biometric identification method based on the photon counting capabilities of the human retina, and the subsequent perception of light. The method offers an unprecedented level of security against malicious attacks. In contrast with existing methods which work within classical physics, we also placed limits on how technologically advanced an impostor has to be in quantum thermometry and quantum magnetometry in order to foil the biometric device by non-invasively monitoring the biometric activity of the device's users. This work opens a venue for exploring quantum optics in a biological context, having both a fundamental scientific interest and the immediate potential for commercial applications in the security industry.

\appendix
\section{}
{\it The probability $\mathbb{P}\big[{\rm see}\big]$ that a coherent light pulse of intensity $I$ and duration $T$ is perceived by a retinal spot of loss parameter $\alpha$ is equal to} $G_K(\alpha I T)$, where $G_K$ is defined in \eqref{GK}.
\newline\newline
{\em Proof:} Photons are incident on the eyball as a Poisson process of intensity $I$. Each incident photon is detected at the retina with probability $\alpha$ independently of others. Hence, photons are detected as a Poisson process of intensity $\alpha I$ (\cite{bertsi}, pp 318). Interval lengths between successive detections are independent exponential random variables with rate $\alpha I$ and the time $T_K$ until the $K$-th detection follows the Erlang distribution with scaling parameter $\alpha I$ and order parameter $K$ (\cite{bertsi}, pp 316). That is, $T_K\sim \frac{1}{\alpha I} E_K$, where $E_K$ is a random variable with cumulative distribution function $G_K$. The pulse will be perceived exactly if $T_K\le T$. Thus,
\[
\mathbb{P}\big[\text{see}\big]=\mathbb{P}\big[T_K\le T\big]=\mathbb{P}\big[E_K\le \alpha I T\big]=G_K(\alpha I T).
\]

\section{}
We will here derive the number of required interrogations for the naive strategy of Sec. IV, namely the estimation of the value of $\alpha$.
\begin{figure*}[t!]
\begin{center}
\includegraphics[width=17.5 cm]{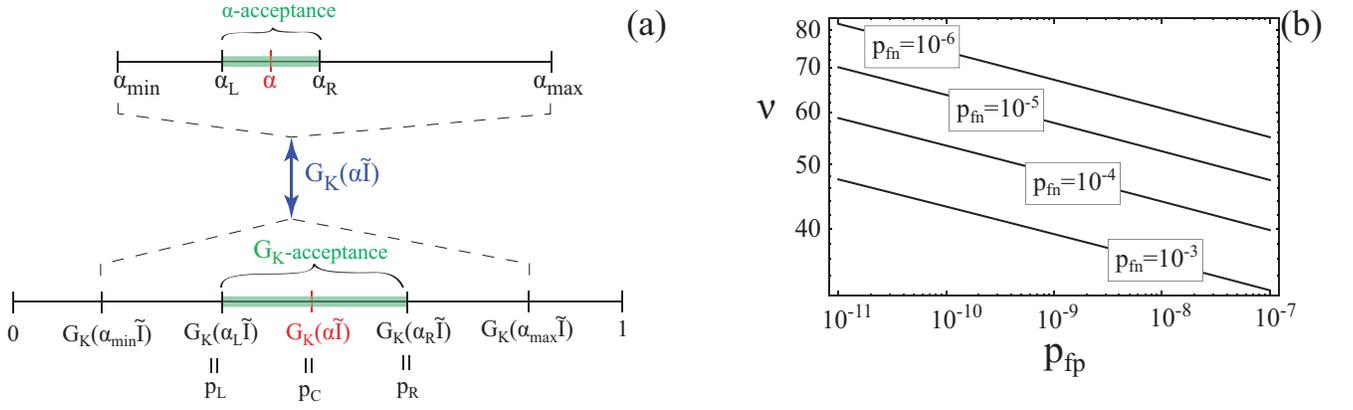}
\caption{(a) The desired level of precision in the estimate of the true $\alpha$ sets a range of acceptable values around the true $\alpha$, given by $\alpha_{L}$ and $\alpha_{R}$, defining the $\alpha$-acceptance region, which lies within the possible $\alpha$ values, ranging from $\alpha_{\rm min}$ to $\alpha_{\rm max}$. The $\alpha$-axis is mapped into the corresponding probability space by the function $G_{K}(x)$, which gives the probability to see a flash when $x$ photons on average are detected by the tested spot of the retina. The inverse map leads from the measured $\mathbb{P}\big[\text{see}\big]$ to $\alpha$. (b) Number of interrogations per spot, $\nu$, required to achieve a false-positive probability $p_{\rm fp}$ for a test involving $\mu$ retinal spots, for various values of the false negative probability $p_{\rm fn}$. For this plot we took $\mu=50$. }
\label{acceptance}
\end{center}
\end{figure*}
For a particular retinal spot being interrogated $\nu$ times, the subject's $j$-th response, where $j=1,...,\nu$, is a Bernoulli random variable $S_j$, taking the values 1=see and 0=don't see. Hence an estimator for $\mathbb{P}[\rm see]$ is 
${1\over \nu}\sum\limits_{j=1}^{\nu}S_{j}$. We will now calculate how many times we have to interrogate a subject with photon pulses (number of pulses $\nu$ for each of the $\mu$ retinal spots) in order to achieve a desired $p_{\rm fp}$ and $p_{\rm fn}$. We will obtain a lower bound on the number of interrogations assuming an impostor, Eve, responds randomly to all $\nu\mu$ interrogations. 

A successful 1-spot test is defined by an acceptance region around Alice's $\alpha$ parameter. In view of the result in Appendix A and the monotonicity of $G_K$ we may recast the test in terms of the estimated $\mathbb{P}[\text{see}]$. Call $p_{C}\equiv G_{K}(\alpha\tilde{I})$ the "correct" probability and consider integers $n_L, n_R$, to be determined later, such that
$p_{L}\equiv \frac{n_L}{\nu}<p_C<\frac{n_R}{\nu}\equiv p_R$. We will consider that the tested subject presenting herself as Alice passes the 1-spot test if $n_L<\sum_{j=1}^{\nu}S_{j}<n_R$. In terms of the $\alpha$-parameter the 1-spot test is passed if the estimated $\alpha$-value lies in the interval $(a_L,a_R)$, where the left and right acceptance limits are defined by $p_{L}\equiv G_{K}(\alpha_{L}\tilde{I})$ and $p_{R}\equiv G_{K}(\alpha_{R}\tilde{I})$ as shown in Fig.\ref{acceptance}a.

Let us now suppose that Eve chooses $p$ uniformly in $(0,1)$ and answers {\em see} with probability $p$, independently for each interrogation. The number of {\em see} answers is then uniformly distributed over the set $[\nu]=\{0,1,\ldots,\nu\}$. Indeed, for $k\in [\nu]$ we have
\begin{align*}
\mathbb{P}\big[\sum_{j=1}^\nu S_j=k\big]&= \int_0^1 {\nu\choose k} p^k(1-p)^{\nu-k}\ dp\\
&={\nu\choose k} \frac{k!(\nu-k)!}{(k+\nu-k+1)!}=\frac{1}{\nu+1}.
\end{align*}
Hence, the probability that Eve successfully passes the 1-spot test is equal to $\frac{n_R-n_L-1}{\nu+1}$. Thus, given the desired false positive probability, $p_{\rm fp}$, we must satisfy the following inequality:
\beq
\frac{n_R-n_L-1}{\nu+1}\leq p_{\rm fp}^{1/\mu}\, .\label{pLR}
\eeq
Next we will obtain a second inequality from the desired false negative probability, $p_{\rm fn}$, which is the probability that Alice fails the test. We will use Lemma 4.7.2 in \cite{Ash} to get a lower bound for the deviation of ${1\over \nu}\sum_{j=1}^{\nu}S_{j}$ from its expectation $p_{C}$:
\begin{align}
\mathbb{P}\Big[\sum_{j=1}^{\nu}{S_{j}}\ge n_{R}\Big]&\geq {1\over \sqrt{8\nu p_R(1-p_R)}}e^{-\nu\mathbb{H}\big(p_{R}\big|p_{C}\big)}\nonumber\\
\mathbb{P}\Big[\sum_{j=1}^{\nu}{S_{j}}\le n_{L}\Big]&\geq {1\over \sqrt{8\nu p_L(1-p_L)}}e^{-\nu\mathbb{H}\big(p_{L}\big|p_{C}\big)}\nonumber
\end{align}
where $\mathbb{H}(x|y)$ is the relative Shannon entropy. Now, let 
\beq
w=\mathbb{P}\Big[\sum_{j=1}^{\nu}S_{j}\ge n_{R}\Big]+\mathbb{P}\Big[\sum_{j=1}^{\nu}S_{j}\le n_{L}\Big]
\eeq
be the probability that Alice fails the 1-spot test. The probability that she fails the whole $\mu$-spot test will then
be the probability that she fails at least one of the $\mu$ tests, and this is equal to $1-(1-w)^\mu$. This should be smaller than the desired false-negative probability $p_{\rm fn}$. Using the preceding estimates and the elementary inequality $4p(1-p)\le 1$ for all $p\in (0,1)$ we arrive at the requirement
\beq
1-\Big[1-{1\over \sqrt{2\nu}}\Big(e^{-\nu {\cal H}\big( p_{R}\big|p_{C}\big)}+e^{-\nu {\cal H}\big(p_{L}\big|p_{C}\big)}\Big)\Big]^{\mu}\leq p_{\rm fn}\label{pfn}
\eeq
To make further progress, suppose $p_{R}-p_{C}\simeq p_{C}-p_{L}$, and use Eq. \eqref{pLR} to find $p_{R}\simeq p_{C}+{1\over 2}p_{\rm fp}^{1/\mu}$ and $p_{L}\simeq p_{C}-{1\over 2}p_{\rm fp}^{1/\mu}$. Using these in Eq. \eqref{pfn} and solving for $\nu$, we obtain the number of interrogations per spot that are required to achieve a desired $p_{\rm fp}$ and $p_{\rm fn}$ for a given number of spots $\mu$. The result is shown in Fig.\ref{acceptance}b. For example, to secure that by responding randomly Eve is positively identified as Alice at most once per 10 billion attempts, and that Alice would fail the test and hence would have to be retested once per ten thousand times, we require slightly more than 50 interrogations for each of 50 retinal spots, i.e. a total of 2500 interrogations.
\section{}
\noindent{\it 
Define $\mu_A(i)=\mathbb{E}_a\left[\ln \left(\frac{Z_A(\alpha_i,S_i)}{Z_E(p,S_i)}\right)\Big|{\cal F}_{i-1}\right]$ and
$\mu_E(i)=\mathbb{E}_E\left[\ln \left(\frac{Z_A(\alpha_i,S_i)}{Z_E(p,S_i)}\right)\Big|{\cal F}_{i-1}\right] $. Then, for all $i\in\mathbb{N}$
\begin{align}
\mu_A(i)&\ge \mathbb{H}\big(q\,\big|\,\frac{1}{2}\big)>0\label{mua}\\
\mu_E(i)&\le \frac{1}{2}\ln\big({4q(1-q)}\big)<0.\label{mue}
\end{align}}

Before we proceed with the proof of this assertion we need to introduce some notation.
Let us denote by $\pi$ the distribution of $G_K(\alpha\tilde{I})$ induced by the random choice of $\alpha$.
This is a probability measure supported on $[0,q]\cup[1-q,1]$, such that for $C\subset [0,1]$,
$\pi(C)=\mathbb{P}\big[G_K(\alpha\tilde{I})\in C\big].$ In particular, we have $p=\mathbb{E}\big[G_K(\alpha\tilde{I})\big]=\int xd\pi(x)$.

Since we randomly target a
high-$\alpha$ or a low-$\alpha$ retinal spot, we must have $\pi([0,q])=\pi([1-q,1])=\frac{1}{2}$.
We will denote by $\pi_L$ ($\pi_H$) the conditional distribution of $G_K(\alpha\tilde{I})$, given that we have selected to target a low (high)-$\alpha$ spot.
That is $\pi_L(C)={2}\pi(C\cap [0,q])$ and $\pi_H(C)={2}\pi(C\cap [1-q,1])$.
Finally, we will denote by $q_L$ (respectively $q_H$) the mean value of these measures, that is
\[
q_L=\int x\, d\pi_L(x)\le q \quad\text{and}\quad q_H=\int x\, d\pi_H(x)\ge 1-q.
\]
Hence,
\[
p=\frac{q_L+q_H}{2}\in (\frac{1-q}{2},\frac{1+q}{2}).
\]

{\em Proof:} We condition first on the value of $\alpha_n$ to get
\begin{align*}
\mu_A(i)&=\mathbb{E}\Big[G_K(\alpha_i\tilde{I})\ln\left(\frac{G_K(\alpha_i\tilde{I})}{p}\right)+\nonumber\\
&\quad + \big(1-G_K(\alpha_i\tilde{I})\big)\ln\left(\frac{1-G_K(\alpha_i\tilde{I})}{1-p}\right)\Big|{\cal F}_{i-1}\Big]\nonumber\\
&=\mathbb{E}\big[ \mathbb{H}\big(G_K(\alpha_i\tilde{I})\big|p\big)\big]\nonumber,
\end{align*}
since $\alpha_i$ is independent of the information available up to time $i-1$. Since $\pi$ is the distribution of $G_K(\alpha\tilde{I})$,
\begin{align*}
\mu_A(i)&=\int \mathbb{H}(x|p)\, d\pi(x).
\end{align*}
Recall that $x\mapsto\mathbb{H}(x|p)$ is decreasing in $[0,p]$ and increasing in $[p,1]$.
We may now split the integral over $x\in [0,q]$ and $x\in [1-q,1]$. Using that $q<p<1-q$,
we further obtain
\begin{align*}
\mu_A(i)&\ge \frac{1}{2}\mathbb{H}(q|p)+\frac{1}{2}\mathbb{H}(1-q|p)\\
&=\mathbb{H}\big(q\,\big|\,\frac{1}{2}\big)+\mathbb{H}(\frac{1}{2}|p)\ge\mathbb{H} \big(q\,\big|\,\frac{1}{2}\big).
\end{align*}
We now turn to \eqref{mue}. 
\begin{align}
\mu_E(i)&=\mathbb{E}_E\Big[p_i\ln\frac{G_K(\alpha_i\tilde{I})}{p}\nonumber\\
&~~~~~~~+(1\!-\!p_i)\ln\frac{1-G_K(\alpha_i\tilde{I})}{1-p}\Big|{\cal F}_{i-1}\Big]\nonumber\\
&\le\max\Big\{\mathbb{E}\big[\ln\frac{G_K(\alpha_i\tilde{I})}{p}\big],\mathbb{E}\big[\ln\frac{1-G_K(\alpha_i\tilde{I})}{1-p}\big]\Big\}\nonumber\\
&=\max\Big\{\int\ln\frac{x}{p}d\pi(x),\int\ln\frac{1-x}{1-p}d\pi(x)\Big\}.
\label{mue1}
\end{align}
We estimate the two terms in \eqref{mue1} by Jensen's inequality.
\begin{align*}
\int \ln\frac{x}{p}\, d\pi(x)&=\frac{1}{2}\int_0^q\ln\frac{x}{p}\, d\pi_L(x)+\frac{1}{2}\int_{1-q}^1\ln\frac{x}{p}\, d\pi_H(x)\nonumber\\
&\le \frac{1}{2}\big(\ln\frac{q_L}{p}+\ln\frac{q_H}{p}\big)=\ln\frac{2\sqrt{q_Lq_H}}{q_L+q_H}.
\end{align*}
It is straightforward that, since $q_L\le q<1-q\le q_H$, the last
expression is maximized when $q_L=q$ and $q_H=1-q$. Hence, 
\[
\int \ln\frac{x}{p}\, d\pi(x)\le \ln\sqrt{4q(1-q)}.
\]
Similarly,
\[
\int \ln\frac{1-x}{1-p}\, d\pi(x)\le \ln\sqrt{4q(1-q)},
\]
and \eqref{mue} follows from \eqref{mue1}. 
\section{}
\noindent{\it The process $\{R_n^{-1}\}_{n\ge 0}$ is a martingale for Alice, and the process process $\{R_n\}_{n\ge 0}$ is a martingale for Eve, regardless of her answering strategy.}\newline\newline
{\em Proof:} We have
\[
\mathbb{E}_A\big[R_{n}^{-1}\big|{\cal F}_{n-1}\big]= R_{n-1}^{-1}\mathbb{E}_A\Big[\frac{Z_E(p,S_n)}{Z_A(\alpha_{n},S_{n})}\big|{\cal F}_{n-1}\Big]\\
\]
If we condition first on the value of $\alpha_n$, the righthand side becomes
\begin{align*}
R_{n-1}^{-1}\mathbb{E}&\Big[G_K(\alpha_{n}\tilde{I})\frac{p}{G_K(\alpha_{n}\tilde{I})}\\
&\quad+\big(1-G_K(\alpha_{n}\tilde{I})\big)\frac{1-p}{1-G_K(\alpha_{n}\tilde{I})}\big|{\cal F}_{n-1}\Big]\\
&=R_{n-1}^{-1}.
\end{align*}
Eve may have her own strategy and there is no reason for her to answer {\em see} with probability $p$, as the test assumes. The probability that she answers {\em see} may change form one interrogation to another, may be random and may even depend on previous answers. However, it may not depend on $\alpha$, as this information is undisclosed. Let us denote by $p_n$ the probability that Eve answers {\em see} to the $n$-th interrogation. We have
\begin{align*}
&\mathbb{E}_E\big[R_{n}\big|{\cal F}_{n-1}\big]= R_{n-1}\mathbb{E}_E\Big[\frac{Z_A(\alpha_{n},S_{n})}{Z_E(p,S_{n})}\big|{\cal F}_{n-1}\Big]\\
&\quad =R_{n\!-\!1}\mathbb{E}_E\big[p_n\frac{G_K(\alpha_{n}\tilde{I})}{p}+(1\!-\!p_n)\frac{1\!-\!G_K(\alpha_{n}\tilde{I})}{1\!-\!p}\big|{\cal F}_{n\!-\!1}\big].
\end{align*}
There are two independent sources of randomness we are integrating in the final equation: Eve's possibly random choice of $p_n$ and the choice of $\alpha_n$.
If we condition first on Eve's answer the expression above becomes
\begin{align*}
&R_{n-1}\mathbb{E}_E\big[p_n\frac{\mathbb{E}\big[G_K(\alpha_n\tilde{I})\big]}{p}\nonumber\\
&~~~~~~~~~~~+(1\!-\!p_n)\frac{1\!-\!\mathbb{E}\big[G_K(\alpha_n\tilde{I})\big]}{1-p}\big|{\cal F}_{n-1}\big]\\
&\qquad=R_{n-1}\mathbb{E}_E\big[p_n\cdot 1+(1\!-\!p_n)\cdot 1\big|{\cal F}_{n-1}\big]=R_{n-1}.
\end{align*}
It is useful to note at this point that the choice $p=\mathbb{E}\big[G_K(\alpha\tilde{I})\big]$ makes the preceding expression independent of $p_n$,  leaving Eve no option to improve her odds by devising a clever strategy.
\section{}
An immediate consequence of \eqref{logodds} and \eqref{mua} is that
$J_n^A=\ln\left(\frac{R_n}{R_0}\right)-n\,\mathbb{H}\big(q\big|\frac{1}{2}\big)$ is a submartingale for Alice. 
We can apply once more the optional stopping theorem to get 
\beq
0\le\mathbb{E}_A\big[J_T^A\big]=\mathbb{E}_A\big[\ln\frac{R_T}{R_0}\big]-\mathbb{H}\big(q\big|\frac{1}{2}\big)\mathbb{E}_A\big[T].
\eeq
By the definition of the stopping time $T$ we must have ${R_{T-1}}<y\, {R_0}$, and since $p\in\big(\frac{1-q}{2},\frac{1+q}{2}\big)$, we must have $\min\{p,1-p\}\ge (1-q)/2$. Hence,
\beq
\frac{R_T}{R_0}=\frac{R_{T-1}}{R_0}\frac{Z_A(\alpha_T,S_T)}{Z_E(p,S_T)}\le \frac{2}{(1-q)p_{\rm fp}}.
\eeq
The last two inequalities together imply that
\begin{equation}
\mathbb{E}_A\big[T\big]\le \frac{\ln\big(\frac{2}{(1-q)p_{\rm fp}}\big)}{\mathbb{H}\big(q\,\big|\,\frac{1}{2})}.
\label{alicetime}
\end{equation}
Likewise, $J_n^E=\ln\left(\frac{R_n}{R_0}\right)-\frac12\ln{4q(1-q)}n$ is a supermartingale for Eve, and the optional stopping theorem gives
\[
\mathbb{E}_E\big[T\big]\le\frac{2\mathbb{E}_E\big[\ln\frac{R_T}{R_0}\big]}{\ln\big(4q(1-q)\big)}\le \frac{2\ln\big(\frac{2q_{\rm min}x}{1+q}\big)}{\ln\big(4q(1-q)\big)},
\]
where $q_{\rm min}=\min\{G_K(\alpha_{\rm min}\tilde{I}), 1-G_K(\alpha_{\rm max}\tilde{I})\}$.


\begin{thebibliography}{0}
\bibitem{note}
It is also possible to get bounds for other statistics of the interrogation time $T$ by considering different (sub/super)-martingales related to the odds ratio $R_n$.
For instance, we can estimate the variance of $T$, or even prove that it has exponentially decaying tails.
\end{thebibliography}
\end{document}